\documentclass[prd,showpacs,superscriptaddress,preprintnumbers,twocolumn,amsmath]{revtex4}
\usepackage[dvips,final]{graphicx}
\usepackage{amssymb}
\usepackage{amsmath}
\usepackage{epsfig}
\usepackage{bm}% bold math
\usepackage{pifont}
\providecommand{\MSbar }{\ensuremath{ \overline{\rm MS} }}
\newcommand{\bit}[1]{\mbox{\boldmath$#1$}}
\newcommand{\vecc}[1]{\mbox{\boldmath $#1$}}

\begin{document}
\preprint{\hbox{RUB-TPII-05/07}}
%\vspace*{-10mm}
%%%%%%%%%%%%%%%%%%%%%%%%%%%%%%%%%%%%%%%%%%%%%%%%%%%%%%%%%%%%%%%%%%%%%%%%

\title{Renormalization, Wilson lines, and transverse-momentum dependent
       parton distribution functions}
\author{I.~O.~Cherednikov}
\email{igorch@theor.jinr.ru}
\affiliation{Institut f\"{u}r Theoretische Physik II,
             Ruhr-Universit\"{a}t Bochum,
             D-44780 Bochum, Germany}
\affiliation{Bogoliubov Laboratory of Theoretical Physics, JINR,
             141980 Dubna, Russia}
\author{N.~G.~Stefanis}
\email{stefanis@tp2.ruhr-uni-bochum.de}
\affiliation{Institut f\"{u}r Theoretische Physik II,
             Ruhr-Universit\"{a}t Bochum,
             D-44780 Bochum, Germany}

\date{\today}
%%%%%%%%%%%%%%%%%%%%%%%%%%%%%%%%%%%%%%%%%%%%%%%%%%%%%%%%%%%%%%%%%%%%%%%

\begin{abstract}
   We perform an analysis of transverse-momentum dependent
   parton distribution functions, making use of their
   renormalization properties in terms of their leading-order
   anomalous dimensions.
   We show that the appropriate Wilson line in the light-cone
   gauge, associated with such quantities, is a cusped one at 
   light-cone infinity.
   To cancel the ensuing cusp anomalous dimension, we include
   in the definition of the transverse-momentum dependent
   parton distribution functions an additional soft counter
   term (gauge link) along that cusped transverse contour.
   We demonstrate that this is tantamount to an ``intrinsic
   (Coulomb) phase'', which accumulates the full gauge history of
   the color-charged particle.

\end{abstract}

\pacs{%
   11.10.Jj, % Asymptotic problems and properties
   12.38.Bx, % Perturbative calculations in QCD
   13.60.Hb, % Total and inclusive cross sections
             % (including deep-inelastic processes)
   13.87.Fh  % Fragmentation into hadrons
}

\maketitle

%%%%%%%%%%%%%%%%%%%%%%%%%%%%%%%%%%%%%%%%%%%%%%%%%%%%%%%%%%%%%%%%%%%%%%%
\section{Introduction}
\label{sec:intro}

A fundamental goal of QCD is to provide an accurate description of
parton distribution functions (PDFs) which contain the nonperturbative
strong dynamics.
While integrated PDFs can be defined in a gauge-invariant way that
is compatible with factorization, ensuring multiplicative
renormalizability and DGLAP evolution, the definition of unintegrated
or, equivalently, transverse-momentum dependent (TMD), parton
distributions, poses severe problems (see, e.g., 
\cite{BR05,Col03,HJ07}):
 (a) Additional, so-called rapidity, divergences \cite{CS81} appear,
 related to lightlike Wilson lines (or the use of the light-cone
 gauge $A^+=0$) \cite{CRS07}, that cannot be taken care of by ordinary
 ultraviolet (UV) renormalization alone.
 In the integrated case these divergences also appear but they
 mutually cancel \cite{CS81,BHKV98}, allowing a probabilistic
 interpretation.
 (b) Moreover, in the light-cone gauge, the result depends on the
 applied pole prescription in the gluon propagator.
 Only with the advanced boundary condition, which sets the transverse
 gauge link to unity, one recovers the results obtained in the Feynman
 gauge \cite{BJY03}.
 (c) The reduction to the integrated case is at least not
 straightforward \cite{JMY04}.
 (d) Universality is in general broken \cite{Col02}, an issue outside
  the scope of our analysis, given that we concentrate on unpolarized
  PDFs only.
  This point will be briefly addressed in the last section. 
Let us discuss these issues in more detail.

The first issue, i.e., the treatment of the rapidity divergences, is on
the focus of our investigation and will be discussed in detail below.
The second question has been addressed by Belitsky, Ji, and Yuan (BJY)
\cite{BJY03} (see also \cite{BMP03} and \cite{JY02}), where a 
transverse gauge link was introduced in order to exhaust the gauge 
freedom of the TMD PDF.
The third problem becomes trivial in our approach because it is
avoided ab initio by the proposed definition of the TMD PDF.
In particular, one may note the
\begin{itemize}
\item Collins-Soper (CS) approach \cite{CS81} (or cutoff method)
      (see also \cite{JMY04}). \\
      These authors were the first to address issue (a) and to propose
      a solution of the problem by adopting either a non-lightlike
      axial gauge or by shifting the integration contour slightly off
      the light cone.
      This, however, entails the introduction of an additional
      rapidity parameter
      $\zeta=(p\cdot n)^{2}/n^{2}$ (with $n^{2}\neq 0$)
      to encode the deviation from the light cone.
      To establish independence from this arbitrary variable, an
      additional evolution equation to the standard one has to be
      fulfilled causing the reduction to the integrated case
      questionable.
      Besides, factorization off the light cone also becomes
      problematic.
\item Collins-Hautmann approach \cite{CH00} (or subtractive method). \\
      These authors suggest another way to circumvent problem (a):
      They restrict themselves to lightlike Wilson lines and remove
      the rapidity divergences by redefining the TMD PDF.
      The principal element in their approach is the introduction
      of a soft counter term that compensates these divergences,
      shown explicitly at the one-loop order and working in the
      Feynman gauge; a fresh look was given by Collins and Metz 
      \cite{CM04}.
      More recently, Hautmann \cite{Hau07} claimed that the reduction
      to the integrated case can also be performed within this method.
\end{itemize}

In our work we will follow another strategy, based on the
renormalization properties of TMD PDFs in terms of their anomalous
dimensions.
The reason is that anomalous dimensions (within perturbative QCD)
encode the key characteristics of Wilson lines in \emph{local} form.
In contrast, gauge contours are \emph{global} objects and, hence,
difficult to handle within a local-field theory framework.
A properly defined TMD PDF should respect collinear factorization.
But this turns out to be in conflict with the gauge link because the
Wilson line contains not only longitudinal gluons that could be
eliminated by imposing the light-cone gauge, it also comprises
transverse ones, with a distinct region of
$\mbox{\boldmath$k_\perp$}$, that are accumulated after the quark
has been struck by the hard current and changes its direction from
$x^+$ to $x^-$.
As a result, one cannot define a TMD PDF by introducing a straight
lightlike line between the quarks (i.e., a ``connector''
\cite{Ste83}).
The reason is that the two quark fields have a separation also 
in the transverse coordinate space and hence the gluons 
originating from this are not collinear to the struck quark 
(they mismatch in the gluon rapidity).
The common assumption to avoid this problem is to use a combined
contour which joins the quarks through light-cone infinity.
Our analysis shows that such a contour cannot be a smooth one, as
usually tacitly assumed, but it has to contain some obstruction 
in the transverse direction (not specified yet) which will inevitably 
contribute to the total anomalous dimension of the TMD PDF.
Therefore, in order to be able to reproduce the well-known result in 
the Feynman gauge, one has to define the TMD PDF in such a way as to 
cancel this unwanted anomalous-dimension term.
To this end, we seek to recast the Wilson line in terms of the
associated anomalous dimensions.
We will show that this can be naturally achieved within a formalism
which inherently respects gauge invariance by using manifestly
gauge-invariant quark fields that account for the whole gauge
``history'' in the sense of Mandelstam \cite{Man62}.
Details will be given elsewhere \cite{CS08}.
The paper is organized as follows.
In the next section we first provide arguments for the necessity to 
insert a transverse gauge link. 
Then we continue with the the calculation of the anomalous dimension 
of the TMD PDF and show that there is a contribution at light-cone 
infinity in the transverse direction that can be associated with a 
cusp.
In the same section we will supply a modified definition of the
TMD PDF that provides the same anomalous dimension as the one
in the Feynman gauge.
Section \ref{sec:Coulomb} deals with the interpretation of the
soft gauge-invariant counter term, introduced in Sec.\
\ref{sec:calc-anom-dim}, as an ``intrinsic Coulomb phase'' in analogy
to the QED case \cite{JS90}.     
Some comments on universality and our conclusions are given in 
Sec.\ \ref{sec:concl}.

\section{Calculation of the anomalous dimension of the TMD PDF}
\label{sec:calc-anom-dim}

\subsection{Transverse gauge link}
\label{subsec:trans-link}

The standard definition of the TMD PDF \cite{CS81}, for a 
quark in a quark distribution supplemented by a transverse link 
\cite{BJY03}, reads
\begin{widetext}
\begin{equation}
\begin{split}
   f_{q/q}(x, \mbox{\boldmath$k_\perp$})
 = {}&
   \frac{1}{2}
   \int \frac{d\xi^- d^2
   \mbox{\boldmath$\xi_\perp$}} {2\pi (2\pi)^2}
   {\rm e}^{- i k^+ \xi^- 
   + i \bit{\scriptstyle k_\perp} \cdot \bit{\scriptstyle \xi_\perp}}
   \left\langle  q(p) |\bar \psi (\xi^-, \xi_\perp)
   [\xi^-, \mbox{\boldmath$\xi_\perp$};
   \infty^-, \mbox{\boldmath$\xi_\perp$}]^\dagger
   [\infty^-, \mbox{\boldmath$\xi_\perp$};
   \infty^-, \mbox{\boldmath$\infty_\perp$}]^\dagger \right.\\
   \quad &
   \left.\gamma^+[\infty^-, \mbox{\boldmath$\infty_\perp$};
   \infty^-, \mbox{\boldmath$0_\perp$}]
   [\infty^-, \mbox{\boldmath$0_\perp$};0^-, \mbox{\boldmath$0_\perp$}]
   \psi (0^-,\mbox{\boldmath$0_\perp$}) |q(p)\right\rangle \
   |_{\xi^+ =0}\ ;\\
[ \infty^-, \mbox{\boldmath$z_\perp$}; z^-, \mbox{\boldmath$z_\perp$}]
\equiv {}&
 {\cal P} \exp \left[
                     i g \int_0^\infty d\tau \ t^{a} \ n_{\mu} 
                     A_{a}^{\mu} (z + n \tau)
               \right], \, 
[ \infty^-, \mbox{\boldmath$\infty_\perp$};
 \infty^-, \mbox{\boldmath$\xi_\perp$}]
\equiv
 {\cal P} \exp \left[
                     i g \int_0^\infty d\tau \ t^{a} \ \mbox{\boldmath$l$} \cdot
                     \mbox{\boldmath$A$}_{a}
                     (\mbox{\boldmath$\xi_\perp$}
                     + \mbox{\boldmath$l$}\tau)
               \right]
\label{eq:tmd_definition}
\end{split}
\end{equation}
\end{widetext}
%Eq (1)
where $\mbox{\boldmath$l_i$}$ represents an arbitrary vector in the
transverse direction and ${\cal P}$ denotes path ordering.
The displayed gauge links 
$[\infty^{-}, \mbox{\boldmath$z_{\perp}$}; 
  z^{-}, \mbox{\boldmath$z_{\perp}$}]
$,
and
$
 [\infty^{-}, \mbox{\boldmath$\infty_{\perp}$};
  \infty^{-}, \mbox{\boldmath$\xi_{\perp}$}]
$
involve gauge contours extending to light-cone infinity in the 
lightlike and in the transverse direction, respectively.
Analogous expressions hold for the other gauge links entering 
(\ref{eq:tmd_definition}).
Belitsky, Ji, and Yuan \cite{BJY03} have shown that the extra 
transverse gauge link is indispensable for the restoration of gauge 
invariance in the light-cone gauge in which the gauge potential does 
not vanish asymptotically.

\begin{figure}
\centering \vspace{0.3cm}
\includegraphics[scale=0.36,angle=90]{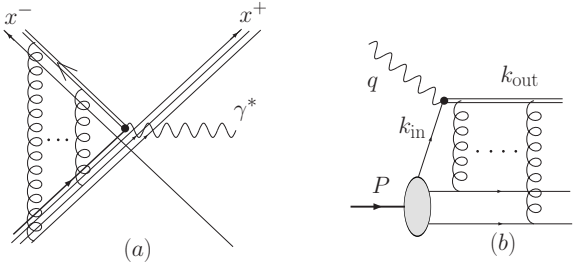}~~
\caption{Spacetime representation $(a)$ and corresponding Feynman graph
         $(b)$ of the collision of a quark with a hard photon in a 
         deeply inelastic process. 
         The struck quark (Wilson line) is denoted by a double line.
         \label{fig:st_wils}}
\end{figure}

The necessity of the additional transverse gauge link in 
Eq.\ (\ref{eq:tmd_definition}) can be most easily understood from the 
point of view of a complete gauge fixing in the axial light-cone gauge. 
Using the spacetime picture of the interaction of a quark, moving fast
in the plus light-cone direction, with the hard spacelike photon, as
depicted in Fig.\ \ref{fig:st_wils}, one can treat the ``classical'' 
current 
\begin{equation}
  j_\mu (y)
  =
  g \int\! d y'_\mu \ \delta^{(4)} (y - y') \  , \quad
  y'_\mu = v_\mu \tau \ , 
\label{eq:current-2}
\end{equation}
%Eq (2)
as a source of the gauge field. 
The gauge field related to such a current has the form
\begin{equation}
  A^\mu (\xi)
  =
  \int\! d^4 y \ D^{\mu\nu}(\xi - y) j_\nu (y)\ \  ,
\label{eq:source1}
\end{equation}
%Eq (3)
where $D^{\mu\nu}$ is the gluon Green's function. 
%%%%%%%%%%%%%%%%%%%%%%%%%%%%%%%%%%%%%%%%%%%%%%%%%%%%%%%%%%%%%%%%%%%%%%%
%                              FIGURE 1
%%%%%%%%%%%%%%%%%%%%%%%%%%%%%%%%%%%%%%%%%%%%%%%%%%%%%%%%%%%%%%%%%%%%%%%
%\begin{widetext}
%\end{widetext}
%%%%%%%%%%%%%%%%%%%%%%%%%%%%%%%%%%%%%%%%%%%%%%%%%%%%%%%%%%%%%%%%%%%%%%%
%%%%%%%%%%%%%%%%%%%%%%%%%%%%%%%%%%%%%%%%%%%%%%%%%%%%%%%%%%%%%%%%%%%%%%%
Appealing to the spacetime structure of this process, illustrated in
Fig.\ \ref{fig:st_wils}(a), we recast the current in the form
\begin{eqnarray}
  j_\mu (y)
& = &
    g \ \delta^{(2)} (\vecc y_\perp) \left[ n^+_\mu \delta (y^-) 
    \int\! \frac{dq^-}{2\pi} \frac{{\rm e}^{-iq^- y^+}}{q^- + i0} 
\right. \\ \nonumber
&  - & \left. n^-_\mu \delta(y^+)  \int\! \frac{dq^+}{2\pi}
    \frac{{\rm e}^{-iq^+ y^-}}{q^+ - i0}\right] \ ,
\label{eq:current-3}
\end{eqnarray}
%Eq (4)
which makes it clear that the first term in this expression
corresponds to a gauge field created by a source moving from minus
infinity to the origin in the plus light-cone direction, before 
being struck by the photon, whereas the second term corresponds to 
a gauge field being created by a source moving from the origin to 
plus infinity along the minus light-cone ray after the collision. 
Then, using the gluon propagator in the light-cone gauge
\begin{eqnarray}
  D^{\mu\nu} (z)
=   
& - \!\!& \! \int\! \frac{d^4 q}{(2\pi)^4} \ 
  \frac{{\rm e}^{- i q z}} {q^2 - \lambda^2 + i0} \nonumber \\
& \times &  
  \left( g^{\mu\nu} - \frac{q^\mu {(n^-)}^{\nu} 
  + q^\nu {(n^-)}^{\mu}}{[q^+]} \right) \ , 
\end{eqnarray}
%Eq (5)
one obtains 
\begin{equation}
  \vecc A_\perp (\infty^-, \vecc \xi_\perp) 
= 
  \frac{g}{4\pi}\, C_\infty \vecc \nabla 
  \ln\left(\lambda |\vecc \xi_\perp|\right) \ ,
\label{eq:ret}
\end{equation}
%Eq (6)
where the numerical constant $C_\infty$ depends on the pole 
prescription applied to regularize the light-cone singularity:
\begin{equation}
   C_\infty
   =
   \left\{
   \begin{array}{ll}
   & \ \ 0  \ , \ {\rm Adv}: \ [q^+] = {q^+ - i 0}    \\
   & - 1 \ , \ {\rm Ret}\, :  \ [q^+] = {q^+ + i 0} \\
   & - \frac{1}{2} \ , \ {\rm PV}\, : \ [q^+]^{-1} 
   = \frac{1}{2}\left (\frac{1}{q^+ + i 0} + \frac{1}{q^+ - i 0}\right)
   \
   \end{array} \right. \ .
   \label{eq:c_inf}
\end{equation}
%Eq (7)
Obviously, the longitudinal components $A^\pm$ vanish. 
On the other hand, the components of the gauge field associated with 
the same source, but in a covariant gauge (labelled by a prime), read
\begin{equation}
   \vecc A'_\perp
= 0 \, ,
\quad
   A'^-
=  0 \, ,
\quad
   A'^+ (\xi)
=
   - \frac{g}{4\pi}\, \delta(\xi^-) 
   \ln\left(\lambda |\vecc \xi_\perp|\right) \ . 
\label{eq:A-prime}
\end{equation}
%Eq (8)
The (singular) gauge transformation, which connects these two field 
representations (i.e., the gauge-field components in the light-cone 
gauge and those in a covariant gauge), is given by 
\begin{equation}
   A_\mu^{\rm LC} = A'_\mu + \partial_\mu \ \phi \, , \quad
   \phi (\xi) =  - \int_{-\infty}^{\xi^-}\!  d\xi'^- A'^+ (\xi'^-) \ .
\label{eq:gauge_trans}
\end{equation}
%Eq (10)
Equation (\ref{eq:gauge_trans}) reflects exactly the gauge freedom 
remaining after fixing the light-cone gauge $A^+ = 0$.
We appreciate that a complete gauge fixing can only be achieved by 
inserting the additional singular gauge transformation 
\begin{eqnarray}
  && U_{\rm sing} (\infty^-, \vecc \xi_{\perp}) 
  = \nonumber \\ 
  && \left[ 1 - i g \int_{-\infty}^{\infty^-}\!
   dz^- A_{\rm source}'^+ (z^-, \vecc z_\perp)
   + O(g^2)
          \right]
          \, .
\label{eq:sing-gauge-transfo}
\end{eqnarray} 
%Eq (11)
which contains the cross-talk effects of the struck parton with the 
light-cone source.
Therefore, the product of two (local) quark field operators in the
completely fixed light-cone gauge (marked below by a wide hat)  
differs from that in a covariant gauge by two phase
factors and attains the form
\begin{eqnarray}
&& \left[ \bar \psi(\xi^{-}, \vecc \xi_{\perp}) \gamma^{+}
   \psi (0^{-}, \vecc 0_{\perp}) \right]_{\widehat{\rm LC}}
=  \nonumber \\ 
&& \bar \psi_{\rm LC} (\xi){\cal P}
   \exp\left[+ ig \int_{\xi_\perp}^{\infty_{\perp}}\!
   d\vecc z_\perp \vecc A_{\rm source}^{\rm LC}
   (\infty^-, \vecc z_\perp) \right] \gamma^+ \
\nonumber \\
&& 
   {\cal P} \exp\left[-  i g\int_{0_\perp}^{\infty_\perp}\!
   d\vecc z_\perp \vecc A_{\rm source}^{\rm LC}
   (\infty^-, \vecc 0_\perp) \right] \psi_{\rm LC} (0) 
\label{eq:lc_perp}
\end{eqnarray}
%Eq (12)
in agreement with Eq.\ (\ref{eq:tmd_definition}). 

\subsection{Anomalous dimensions}
\label{subsec:anom-dim}

Within the CS approach, where $n^2\neq 0$, the anomalous dimension
associated with $f_{q/q}(x, \vecc k_{\perp})$ is
\begin{eqnarray}
    \gamma_{\rm CS}
& = &
    \frac{1}{2} \
    \mu \frac{d}{d \mu}
    \ln Z_{f} (\mu, \alpha_s; \epsilon) 
=
    \frac{3}{4} \frac{\alpha_s}{\pi} C_{\rm F} + O (\alpha_s^2)
\nonumber\\
& = &
    \gamma_{\rm smooth}\ ,
\label{eq:gamma-smooth}
\end{eqnarray}
%Eq (12)
where $Z_f$ is the renormalization constant of
$f_{q/q}(x, \vecc k_{\perp})$ in the {\MSbar}~scheme.
As long as one assumes that the deformation of the Wilson line in the
transverse direction off the light cone preserves the smoothness of the 
gauge contour, the associated anomalous dimension is only
due to the endpoints and, therefore, equals that of the connector
insertion \cite{Ste83}.
Hence, as far as the renormalization of the pure gauge link with a 
finite contour is concerned, the straight lightlike line is enough to 
supply its anomalous dimension because other contour characteristics, 
e.g., its length, are irrelevant.   

\begin{figure}
\centering \vspace{0.3cm}
\includegraphics[scale=0.36,angle=90]{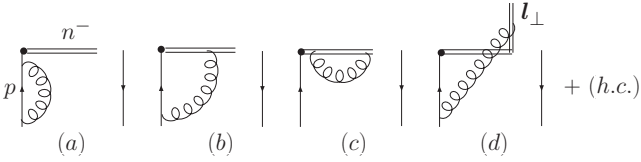}~~
\caption{One-loop gluon contributions to the UV-divergences of the
         TMD PDF.
         Double lines denote gauge links.
         Diagrams (b) and (c) are absent in the light-cone gauge.
\label{fig:se_gluon}}
\end{figure}

In general, in renormalizing the distribution $f_{q/q}$ of a quark in
a quark, one faces UV divergences stemming from the momentum
integration that can be renormalized in the usual way.
But, using the light-cone gauge $n^2=0$, extra UV divergences
contribute due to the additional pole of the gluon propagator,
as already mentioned.
We calculate the UV divergences of the one-loop diagrams, shown in
Fig.\ \ref{fig:se_gluon}, which contribute to
$f_{q/q}(x, \mbox{\boldmath$k_\perp$})$
in the light-cone gauge
$(A \cdot n^-) = 0, \ {(n^-)}^2 = 0$,
by using dimensional regularization.
The poles $1/q^+$ of the gluon propagator
\begin{equation}
   D_{\mu\nu}^{\rm LC} (q)
=
   \frac{1}{q^2} \Big( g_{\mu\nu}
  -\frac{q_\mu n^-_\nu + q_\nu n^-_\mu}{[q^+]}\Big) \ ,
\end{equation}
%Eq (13)
are regularized according to 
\begin{equation}
  \frac{1}{[q^+]}
= 
  \frac{1}{q^+ \pm i \Delta}\, .
\label{eq:Delta}
\end{equation}
%Eq (14).
In what follows, we keep $\Delta$ small but finite.

%%%%%%%%%%%%%%%%%%%%%%%%%%%%%%%%%%%%%%%%%%%%%%%%%%%%%%%%%%%%%%%%%%%%%%%
%                              FIGURE 2
%%%%%%%%%%%%%%%%%%%%%%%%%%%%%%%%%%%%%%%%%%%%%%%%%%%%%%%%%%%%%%%%%%%%%%%
%\begin{widetext}
%\end{widetext}
%%%%%%%%%%%%%%%%%%%%%%%%%%%%%%%%%%%%%%%%%%%%%%%%%%%%%%%%%%%%%%%%%%%%%%%
%%%%%%%%%%%%%%%%%%%%%%%%%%%%%%%%%%%%%%%%%%%%%%%%%%%%%%%%%%%%%%%%%%%%%%%

The UV divergent part of diagrams (a) and (d) in Fig.\ 
\ref{fig:se_gluon} receives contributions owing to the $p^+$-dependent 
term
\begin{equation}
    \Sigma_{\rm LC}^{\rm UV} (\alpha_s, \epsilon) 
=
    \frac{\alpha_s}{\pi}C_{\rm F} 2\! \left[ \frac{1}{\epsilon }
    \left( \frac{3}{4} 
  + \ln \frac{\Delta}{p^+} \right) - \gamma_E + \ln 4\pi \right]   
\label{eq:gamma_1}
\end{equation}
%Eq (15)
in addition to those originating from the standard UV renormalization.
In deriving expression (\ref{eq:gamma_1}), we find that the
contribution associated with the transverse gauge link at infinity
(diagram Fig.\ 1(d)) exactly cancels against the term entailed by the
adopted pole prescription in the gluon propagator.
This confirms the previous results by BJY and establishes the
dependence of the result on local quantities only.
Therefore, the corresponding anomalous dimension is given by
\begin{equation}
  \gamma_{\rm LC}
=
  \frac{\alpha_s}{\pi}C_{\rm F}\Bigg( \frac{3}{4}
  + \ln \frac{\Delta}{p^+} \Bigg)
=
  \gamma_{\rm smooth} - \delta \gamma \ .
\label{eq:gamma_2}
\end{equation}
%Eq (16)
The difference between
$\gamma_{\rm smooth}$ and $\gamma_{\rm LC}$ is exactly that term
induced by the additional divergence which has to be compensated
by a suitable redefinition of the TMD PDF.
Note that 
$p^+ = (p \cdot n^-) \sim \cosh \chi$ 
defines, in fact, an angle $\chi$ between the direction of the 
quark momentum $p_\mu$ and the lightlike vector $n^-$.
In the large $\chi$ limit,
$\ln p^+ \to \chi , \ \chi \to \infty$.
Thus, we can conclude that the ``defect'' of the anomalous dimension,
$\delta \gamma$, can be identified with the well-known cusp anomalous
dimension \cite{KR87}
\begin{equation}
\begin{split}
   & \gamma_{\rm cusp} (\alpha_s, \chi)
= \frac{\alpha_s}{\pi}C_{\rm F} \ (\chi \coth \chi - 1 ) \ , \\
& \frac{d}{d \ln p^+} \ \delta \gamma
= \lim_{\chi \to \infty}
  \frac{d}{d \chi} \gamma_{\rm cusp} (\alpha_s, \chi)
= \frac{\alpha_s}{\pi}C_{\rm F} \ .
\end{split}
\end{equation}
%Eq (17)
This provides formal support for our previous statement concerning the
appropriate choice of the Wilson line in the definition of the TMD PDF.

%%%%%%%%%%%%%%%%%%%%%%%%%%%%%%%%%%%%%%%%%%%%%%%%%%%%%%%%%%%%%%%%%%%%%%%
%                              FIGURE 3
%%%%%%%%%%%%%%%%%%%%%%%%%%%%%%%%%%%%%%%%%%%%%%%%%%%%%%%%%%%%%%%%%%%%%%%
%\begin{widetext}
 \begin{figure}
\centering \vspace{0.6cm}
\includegraphics[scale=0.5]{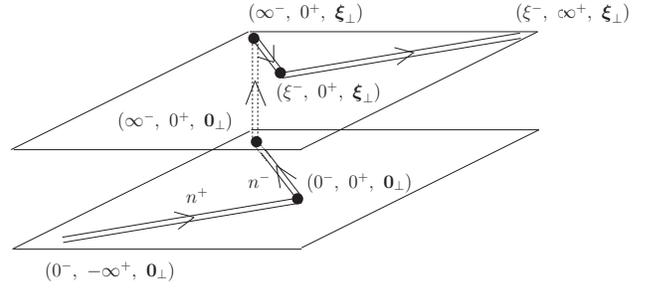}~~
\caption{The integration contour associated with the additional soft
         counter term.
\label{fig:contour}}
\end{figure}
%\end{widetext}
%%%%%%%%%%%%%%%%%%%%%%%%%%%%%%%%%%%%%%%%%%%%%%%%%%%%%%%%%%%%%%%%%%%%%%%
%%%%%%%%%%%%%%%%%%%%%%%%%%%%%%%%%%%%%%%%%%%%%%%%%%%%%%%%%%%%%%%%%%%%%%%

As one knows from the renormalization of contour-dependent composite
operators in QCD (see \cite{Pol80} and also \cite{KR87,CD80,Aoy81} and 
earlier references cited therein), the standard UV renormalization 
procedure has to be generalized in order to be able to subtract 
angle-dependent singularities stemming from obstructions, like cusps 
or self-intersections.
Having recourse to these techniques, we compute the extra
renormalization constant associated with the soft counter term
\cite{CH00} and show that it can be expressed in terms of a
vacuum expectation value of a specific gauge link.
In order to cancel the anomalous dimension defect
$\delta \gamma$,
we introduce the counter term
\begin{equation}
  R
\equiv
 \Phi (p^+, n^- | 0) \Phi^\dagger (p^+, n^- | \xi) \ ,
\label{eq:soft_factor_1}
\end{equation}
%Eq (18)
where
\begin{equation}
  \Phi (p^+, n^- | \xi )
 =
  \left\langle 0
  \left| {\cal P} \exp\Big[ig \int_{\Gamma_{\rm cusp}}d\zeta^\mu
  \ t^a A^a_\mu (\xi + \zeta)\Big]
  \right|0
  \right\rangle
\label{eq:soft_definition}
\end{equation}
%Eq (19)
and evaluate it along the non-smooth, off-the-light-cone integration 
contour
$\Gamma_{\rm cusp}$,
defined by
\begin{eqnarray}
\begin{split}
& \Gamma_{\rm cusp} : \ \ \zeta_\mu
=
  \{ [p_\mu^{+}s \ , \ - \infty < s < 0] \  \\
& \cup \ [n_\mu^-  s' \ ,
  \ 0 < s' < \infty] \ \cup \
  [ \mbox{\boldmath$l_\perp$} \tau , \, \ 0 < \tau < \infty ] \}
\label{eq:gpm}
\end{split}
\end{eqnarray}
%Eq (20)
$\rule{0in}{2.6ex}$
with $n_\mu^-$ being the minus light-cone vector, illustrated in
Fig.\ \ref{fig:contour}.

\begin{figure}
\centering \vspace{-0.2cm}
\includegraphics[scale=0.36,angle=90]{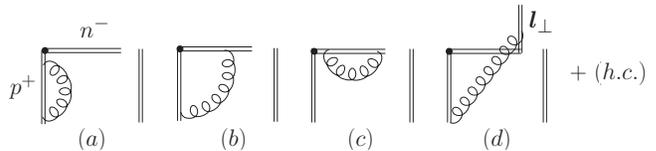}~~
\caption{Virtual gluon contributions to the UV-divergences of the soft
         counter term (in analogy to Fig.\ \ref{fig:se_gluon}).
\label{fig:soft_gluon}}
\end{figure}

The one-loop gluon virtual corrections, contributing to the UV
divergences of $R$, are shown in Fig.\ \ref{fig:soft_gluon}.
For the UV divergent term we obtain
\begin{equation}
  \Sigma_{R}^{\rm UV}
=
  - \frac{ \alpha_s}{\pi} C_{\rm F} \ 2 \left(  \frac{1}{\epsilon} \
  \ln \frac{\Delta}{p^+} - \gamma_E + \ln 4 \pi \right)
\end{equation}
%Eq (21)
and observe that this expression is equal, but with opposite sign,
to the unwanted term in the UV singularity, related to the cusped
contour, calculated before.

Hence, it is reasonable to redefine the conventional TMD PDF and
absorb the soft counter term in its definition.
Then we have 
\begin{widetext}
\begin{equation}
\begin{split}
   f_{q/q}^{\rm mod}(x, \mbox{\boldmath$k_\perp$})
 = {}&
  \frac{1}{2}
   \int \frac{d\xi^- d^2
   \mbox{\boldmath$\xi_\perp$}}{2\pi (2\pi)^2}
   {\rm e}^{- i 
   \bit{\scriptstyle k_\perp} \cdot \bit{\scriptstyle \xi_\perp}}   
   \left\langle  q(p) |\bar \psi (\xi^-, \xi_\perp)
   [\xi^-, \mbox{\boldmath$\xi_\perp$};
   \infty^-, \mbox{\boldmath$\xi_\perp$}]^\dagger
   [\infty^-, \mbox{\boldmath$\xi_\perp$};
   \infty^-, \mbox{\boldmath$\infty_\perp$}]^\dagger \right.\\
   \quad &
   \left.\gamma^+[\infty^-, \mbox{\boldmath$\infty_\perp$};
   \infty^-, \mbox{\boldmath$0_\perp$}]
   [\infty^-, \mbox{\boldmath$0_\perp$}; 0^-,\mbox{\boldmath$0_\perp$}]
   \psi (0^-,\mbox{\boldmath$0_\perp$}) |q(p)\right\rangle \
   \cdot
   \Big[ \Phi(p^+, n^- | 0^-, \mbox{\boldmath$0_\perp$})
   \Phi^\dagger (p^+, n^- | \xi^-, \mbox{\boldmath$\xi_\perp$})
   \Big] \, ,
\label{eq:tmd_re-definition}
\end{split}
\end{equation}
\end{widetext}
%Eq (22)
which is one of the main results of our work.
For the renormalization of
\begin{equation}
  f_{\rm ren}^{\rm mod}(x, \mbox{\boldmath$k_\perp$})
=
  Z_{f}^{\rm mod} (\alpha_s, \epsilon)
  f^{\rm mod} (x, \mbox{\boldmath$k_\perp$}, \epsilon)
\label{eq:renorm-f}
\end{equation}
%Eq (23)
the standard UV renormalization is sufficient.
It yields the following renormalization constant
\begin{eqnarray}
   Z_{f}^{\rm mod}
& = &
  1 + \frac{ \alpha_s}{4\pi}\, C_{\rm F} \ \frac{2}{\epsilon}
  \left(- 3 - 4 \ln \frac{\Delta}{p^+}
  + 4 \ln \frac{\Delta}{p^+} \right)\nonumber \\
& = &
  1 - \frac{ 3\alpha_s}{4\pi}\, C_{\rm F} \ \frac{2}{\epsilon} \ ,
\end{eqnarray}
%Eq (24)
which in turn gives rise to the anomalous dimension
\begin{equation}
   \gamma_{f}^{\rm mod}
=
  \frac{1}{2} \mu \frac{d}{d\mu}
  \ln Z_{f}^{\rm mod} (\mu , \alpha_s, \epsilon)
= \frac{3}{4} \frac{\alpha_s}{\pi}\, C_{\rm F} + O(\alpha_s^2)\, .
\end{equation}
%Eq (25)
It is obvious that (at least at the one-loop order) this expression
coincides with $\gamma_{\rm smooth}$ given by Eq.\
(\ref{eq:gamma-smooth}).

\section{Intrinsic Coulomb phase}
\label{sec:Coulomb}

The physical meaning of the introduced soft counter term can be
described as follows.
Appealing to the exponentiation theorem for non-Abelian path-ordered
exponentials \cite{KR87}, the vacuum average
(\ref{eq:soft_definition}) can be recast in the form
\begin{equation}
  \Phi (u, n^-)
=
  \exp \left[\sum_{n=1}^{\infty} \alpha_s^n \Phi_n (u, n^-) \right] \ ,
\end{equation}
%Eq (26)
where the functions $\Phi_n$ have, in general, a complicated
structure.
Nevertheless, the leading term in this series, $\Phi_1$, is just a
non-Abelian generalization of the Abelian expression
\begin{equation}
  \Phi_1 (u, n^-)
= - 4 \pi C_{\rm F} \
  \int_{\Gamma_{\rm cusp}}\! dx_\mu dy_\nu \
  \theta (x-y)\ D^{\mu\nu} (x-y) \ .
\label{eq:phase_1}
\end{equation}
%Eq (27)
%%%%%%%%%%%%%%%%%%%%%%%%%%%%%%%%%%%%%%%%%%%%%%%%%%%%%%%%%%%%%%%%%%%%%%%
%                              FIGURE 4
%%%%%%%%%%%%%%%%%%%%%%%%%%%%%%%%%%%%%%%%%%%%%%%%%%%%%%%%%%%%%%%%%%%%%%%
%\begin{widetext}
%\end{widetext}
%%%%%%%%%%%%%%%%%%%%%%%%%%%%%%%%%%%%%%%%%%%%%%%%%%%%%%%%%%%%%%%%%%%%%%%
%%%%%%%%%%%%%%%%%%%%%%%%%%%%%%%%%%%%%%%%%%%%%%%%%%%%%%%%%%%%%%%%%%%%%%%
By virtue of the current
\begin{equation} 
  j_\nu^b (z)
=
  t^b\ v_\nu\ 
  \int_{\Gamma_{\rm cusp}} d\tau\ \delta^{(4)}(z - v\tau),
\label{eq:current}
\end{equation}
%Eq (28)
evaluated along the contour
$\Gamma_{\rm cusp}$ (cf.\ Eq. (\ref{eq:gpm})) and
where the velocity $v_\nu$ equals either $u_\nu$, $n^-$, or
$\mbox{\boldmath$l_\perp$}$ (depending on the segment of the contour
along which the integration is performed), one can rewrite
(\ref{eq:phase_1}) as follows
\begin{equation}
 \Phi_{1}(u, n^-) = \\
    -   t^a 4\pi
\int_{\Gamma_{\rm cusp}}\! dx_\mu  \int d^4 z 
 \delta^{ab} D^{\mu\nu} (x-z) j_\nu^b (z) \, .
  \label{eq:coul}
\end{equation}
%Eq (29)

This result proves that the additional soft counter term $R$ can be 
treated within Mandelstam's manifestly gauge-invariant formalism and 
appears there as an ``intrinsic Coulomb phase'' \cite{JS90} originating 
from the long-range interactions of a colored quark, created initially 
at the ``point'' $-\infty^+$ together with its oppositely color-charged
counterpart, then travelling along the plus light-cone ray to the
origin, where it experiences a hard collision with the off-shell
photon, subsequently changing its route and venturing along the minus
ray to $+\infty^-$.
Within such a context, the soft counter term can be conceived of as
that part of the TMD PDF which accumulates the residual effects of the
primordial separation of two oppositely color-charged particles,
created at light-cone infinity and being unrelated to the existence of
external color sources.

\section{Discussion and Conclusions}
\label{sec:concl}

The study presented above was performed for the semi-inclusive DIS 
(SIDIS). 
Before we conclude, it is appropriate to make some comments concerning 
the Drell-Yan lepton-pair production. 
In this case, it is known \cite{Col02} that the direction of the 
integration contours in the gauge links should be reversed. 
In the light-cone gauge, this corresponds  to a change of sign of the 
additional regulator $\Delta$ (cf.\ Eq.\ (\ref{eq:Delta})) 
\begin{equation} 
    \Delta_{\rm SIDIS} = 
    - \Delta_{\rm DY}  \ . 
\label{eq:sidis_dy}  
\end{equation}    
%Eq (30)
In the non-polarized case, this affects only the imaginary parts, and, 
therefore, it does not contribute to the final expressions. 
In other words, the UV anomalous dimension of the non-polarized TMD 
PDF's is universal as regards the SIDIS and the DY processes. 
This, however, may not be true for the spin-dependent TMD PDF's, 
since in that case the imaginary parts play a crucial role and, thus, 
a sign change (expressed in (\ref{eq:sidis_dy})) might indeed affect 
the renormalization-group properties and the corresponding evolution 
equations. 
These issues will be considered elsewhere.     

Let us summarize the cornerstones of our work:\\
(i) We performed an analysis of TMD PDFs based on anomalous dimensions
that encapsulate the relevant Wilson-line characteristics in local
form.\\
(ii) We showed by explicit calculation at the one-loop level that 
the appropriate Wilson contour in the light-cone gauge is a cusped 
one, contributing an angle-dependent anomalous dimension to the TMD 
PDF, that has to be compensated in order to render it compatible 
with the collinear factorization.
The validation of this cancellation in next-to-leading order is 
currently in progress.\\
iii) We outlined how this new contribution can be included in the
definition of the TMD PDF by means of a soft counter term, as 
proposed by Collins \cite{Col02}.
We found that this new term can be written as an ``intrinsic Coulomb 
phase'' that keeps track of the full gauge history of the colored 
quarks \cite{JS90}.

This phase may be given the following interpretation:
Before the quark is being struck it is escorted only by longitudinal
gluons that can be formally eliminated by imposing the light-cone
gauge.
However, when it leaves the (hard) interaction region, it is not lying
on the minus light-cone direction and exchanges soft transverse gluons
with the quark spectator.
Hence, one cannot trivialize the interaction of the struck quark with 
the gauge field by imposing a single gauge choice on a lightlike ray,
because the struck quark enters and leaves the hard-interaction vertex 
with different four-velocities, as it becomes evident from Eq.\ 
(\ref{eq:current}). 
This complies with the interpretation given by Belitsky, Ji, and Yuan 
\cite{BJY03} (see also \cite{JY02}) in terms of final-state 
interactions of the struck quark with the gluon field of the target 
spectators.
As long as one disregards polarization effects, the direction of the
Wilson line (expressed by means of the $i\epsilon$ prescription in the 
gluon propagator), appears only in intermediate steps of the calculation
and cancels at the end because it is only a phase.
In conclusion, our analysis may lead to a deeper insight of the 
dynamics of TMD PDFs and have wide-range phenomenological applications.

%%%%%%%%%%%%%%%%%%%%%%%%%%%%%%%%%%%%%%%%%%%%%%%%%%%%%%%%%%%%%%%%%%%%%%%
\acknowledgments
We thank A.P.\ Bakulev, N.A.\ Kivel, A.I.\ Karanikas,
S.V.\ Mikhailov, P.V.\ Pobylitsa, and O.V.\ Teryaev for
discussions.
I.O.C. is supported by the Alexander von Humboldt Foundation.
This work was supported in part by the Deutsche Forschungsgemeinschaft
under grant 436 RUS 113/881/0, Russian Federation President's grant
1450-2003-2, and the Heisenberg--Landau Program 2007 and 2008.

%%%%%%%%%%%%%%%%%%%%%%%%%%%%%%%%%%%%%%%%%%%%%%%%%%%%%%%%%%%%%%%%%%%%%%%

%%%%%%%%%%%%%%%%%%%%%%%%%%%%%%%%%%%%%%%%%%%%%%%%%%%%%%%%%%%%%%%%%%%%%%%
\end{document}